\documentclass[aps,showpacs,superscriptaddress,twocolumn,prl]{revtex4}
\usepackage{amssymb}
\usepackage{natbib}
\usepackage{amsmath}
\usepackage{amsfonts}
\usepackage{graphicx}
\usepackage{mathrsfs}
\usepackage{dcolumn}
\usepackage{bm}
\usepackage{color}
\usepackage{cancel}
\usepackage{mathbbol}
\usepackage{epsfig}
\usepackage{units}
\usepackage{esint}
\usepackage{soul}

\definecolor{rot}{rgb}{0.75,0.05,0.25}
\definecolor{hellgrau}{gray}{0.5}
\definecolor{blau}{rgb}{0,0,0.7}

\def\Tr{\mbox{Tr}}

\begin{document}

\title{Dissipation, Correlation and Lags in Heat Engines}
\author{Michele Campisi}
\affiliation{NEST, Scuola Normale Superiore \& Istituto Nanoscienze-CNR, I-56126 Pisa, Italy}
\author{Rosario Fazio}
\affiliation{ICTP, Strada Costiera 11, 34151 Trieste, Italy}
\affiliation{NEST, Scuola Normale Superiore \& Istituto Nanoscienze-CNR, I-56126 Pisa, Italy}
\date{\today }

\begin{abstract}
By modelling heat engines as driven multi-partite system we show that their dissipation can be expressed in terms of  the lag (relative entropy) between the perturbed state of each partition and their equilibrium state, and the correlations that build up among the partitions.  We illustrate the rich interplay 
between  correlations and lags with a two-qubit device driven by a quantum gate.
\end{abstract}

\pacs{
}

\maketitle

\section{Introduction}
Quantum thermodynamics, the discipline that studies the impact of the laws of quantum mechanics on the transduction of work into heat 
(and vice-versa) in the microscopic domain, is currently undergoing a fast and intense development~\cite{Benenti13arXiv13114430,Kosloff14ARPC65,
Vinjanampathy15arXiv150806099,Goold15arXiv150507835,GelbwaserKlimovsky15AAMOP64,Millen16NJP18}. One of its main objectives concerns 
the understanding of quantum  thermal machines: Their underlying functionalities, mode of operation, bounds that limit their performance, as well 
as the conception and experimental realisation thereof. 

\begin{figure}[b]
\includegraphics[width=\linewidth]{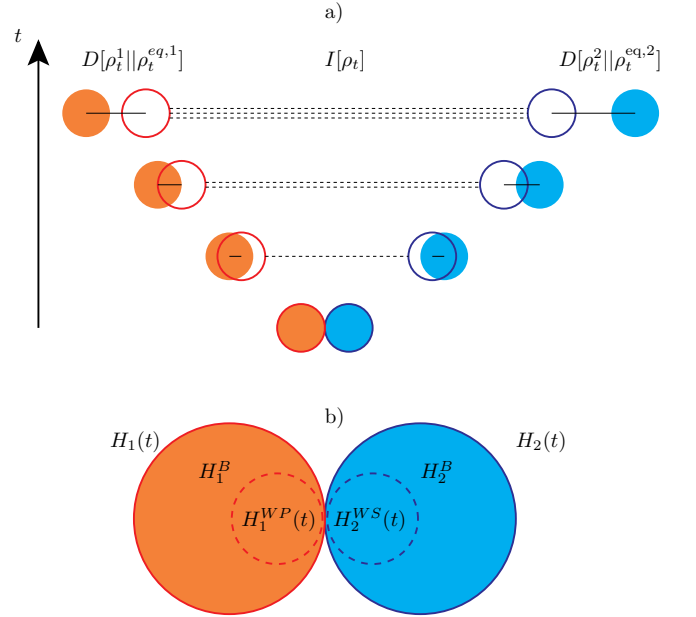}
\caption{(Color online) Panel a): Sources of dissipation in a multipartite system. During the drive the state of each subsystem (empty disks) lags 
(solid line) behind the reference equilibrium (filled disks), and develops correlations (dashed lines) with the other subsystem. Panel b): A heat 
engine working with two reservoirs is modelled as a bipartite system starting in the factorised initial state (\ref{eq:eq-initial1}).}
\label{fig:1}
\end{figure}

Among the recent advancements in the field of non-equilibrium (classical and quantum) thermodynamics~\cite{Esposito09RMP81,Campisi11RMP83},  
the fluctuation relation for heat engines \cite{Sinitsyn11JPA4,Campisi14JPA47,Campisi15NJP17} establishes the microscopic conditions under which 
a generic heat engine cannot have an efficiency overcoming the Carnot efficiency. 
In its essence a heat engine can be imagined composed by a working substance and two (or more) thermal reservoirs. The whole system can be 
modelled as a driven multi-partite system starting in a factorised state~\cite{Campisi14JPA47,Campisi15NJP17}
\begin{align}
\rho_0 = \frac{e^{-\beta_1 H_1}}{Z_1} \otimes \frac{e^{-\beta_2 H_2}}{Z_2} \otimes  \cdots \otimes \frac{e^{-\beta_N H_N}}{Z_N} \, .
\label{eq:eq-initial1}
\end{align}
Here $H_i$ includes both the Hamiltonian of bath $i$, $H^B_i$ and possibly the Hamiltonian $H^{WS}_i$ of one sub-part (part $i$) of the working substance 
(as well as their mutual interactions), see Fig. \ref{fig:1}, panel b (although we use a quantum notation, the theory developed in this paper applies unaltered 
in classical mechanics).  
Machines achieving efficiencies larger than Carnot's can be conceived. This however can only happen when the above modelling does not apply, e.g., 
when the initial state contains non-factorisable correlation terms, coherences~\cite{Scully03Science299}, or if it contains non-thermal states, e.g., squeezed 
thermal states~\cite{Rossnagel14PRL112}.  

Here we continue the investigation of heat engines along the lines set in Refs. \cite{Campisi14JPA47,Campisi15NJP17} and study the sources of 
dissipation in a heat engine by analysing the non-equilibrium behaviour of a generic driven multi-partite system.
In the case of a mono-partite system prepared in a thermal state one of the cornerstones of modern non-equilibrium thermodynamics  establishes that the 
total dissipation during the driving can be quantified by means of an information theoretic quantity, namely the relative entropy $D[\rho_t||\rho_t^\text{eq}]=
\Tr (\rho_t \ln \rho_t-\rho_t\ln \rho_t^\text{eq})$ between the off-equilibrium state $\rho_t$, i.e. the time evolved of $\rho_0^\text{eq}$, and the corresponding 
equilibrium $\rho_t^\text{eq}=e^{-\beta H(\lambda_t)}/Z(\lambda_t,\beta)$~\cite{Kawai07PRL98,Vaikuntanathan09EPL87,Deffner10PRL105}:
\begin{align}
\beta W_\text{diss} = D[\rho_t||\rho_t^\text{eq}] \label{eq:Wdiss}
\end{align}
Here, $H(\lambda_t)$ is the systems Hamiltonian depending on a time-dependent parameter $\lambda_t$, $W_\text{diss}=\langle W\rangle-\Delta F$, 
is the average dissipate work with $\langle W\rangle$ the average work and $\Delta F$, the free energy difference between the initial state $\rho_0^\text{eq}$ 
and the reference thermal state $\rho_t^\text{eq}$ at time $t$.  This relation has been useful for analysing and understanding fundamental questions 
such as the Landauer principle~\cite{Kawai07PRL98,Berut12Nature483}. 

Here we generalise Eq. (\ref{eq:Wdiss}) to the case of a multipartite system in contact with many baths. The corresponding expression, our main result, reads
\begin{align}
\sum_i \beta_i W_\text{diss}^i = \sum_i D[\rho^i_t || \rho_t^\text{i,eq}] + I[\rho_t]
\label{eq:sumbetaWdiss}
\end{align}
where $W_\text{diss}^i$ is the work dissipated in each part, and $D[\rho^i_t || \rho_0^\text{i,eq}]$ denotes the relative entropy between the reduced 
density matrix of part $i$, $\rho_t ^i$, and its the reference equilibrium state $\rho_t^\text{i,eq}$. Most remarkably, besides the lags $D[\rho^i_t || \rho_t^\text{i,eq}]$, 
another term $I[\rho_t]$ occurs in Eq. (\ref{eq:sumbetaWdiss}) which is the total correlation among all the parts composing the system, (that is the mutual 
information in the case of bi-partite system). In a multipartite system there are two sources of dissipation: The lag of each subsystem with respect to its equilibrium 
and the correlation which builds up among the parts. Eq. (\ref{eq:sumbetaWdiss}) is illustrated in Fig. \ref{fig:1}, panel a). That correlation is a source of dissipation 
was established previously in Ref. \cite{Esposito10NJP12}, which focussed on an open system, starting in a generic state, and identified entropy production 
with the system bath correlations. The present work thus unifies the result of Esposito {\it et al} with Eq.(\ref{eq:Wdiss}). 

In the specific case when the driven multi-partite system is a heat engine working with $N$ reservoirs at inverse temperatures $\beta_1\leq 
\beta_2 \dots \leq \beta_N$, Eq. (\ref{eq:sumbetaWdiss}) implies
\begin{align}
\beta_N W_\text{out} \delta \eta \geq  \sum_i D[\rho^i_\tau || \rho_0^\text{i,eq}] +  I[\rho_\tau]
\label{eq:complementarity}
\end{align}
where $\delta \eta = \eta^C/\eta -1 $ is the relative deviation of the heat engine's efficiency from Carnot's efficiency $\eta^C=1-\beta_1/\beta_N$ 
and $W_\text{out}\geq0$ is the heat engine's output work. 
Eq. (\ref{eq:complementarity}) has far-reaching consequences for the microscopic theory of heat engines (i) no quantum (nor classical) thermal machine 
can overcome Carnot efficiency. This is a consequence of the fact that the r.h.s of Eq. (\ref{eq:complementarity}) is non-negative. (ii) Carnot efficiency 
can only be reached at zero work output. We shall detail both points below.

We further illustrate the main result of this work by considering an experimentally feasible quantum engine based on a double qubit, undergoing 
gate operations~\cite{Campisi15NJP17}. This example also helps shading light onto the question of what is the role of correlations in quantum 
thermodynamics, a topic of great current interest~\cite{Correa14SCIREP4,Apollaro15PSCT165,Huber15NJP17}. As we shall see depending on whether 
one looks at efficiency or work, correlations can be either an obstacle or a help.

\section{Dissipation in driven multi-partite systems}
Consider a driven multipartite system:
\begin{align}
H(t)= H_1(t) + H_2(t) + \dots H_N{(t)} + V(t)
\end{align}
$H_i(t)$ denotes the time dependent Hamiltonian of subsystem $i$, while $V(t)$ denotes a time dependent coupling among the subsystems.
The driving consists in the temporal sequence $t$ between $t=0$ and $t=\tau$. The coupling $V$ is assumed to vanish for all times $t \notin (0,\tau)$, so that the driving moves the Hamiltonian from 
$H(0)= \sum_i H_i(0)$ to $H(\tau)= \sum_i H_i(\tau)$.
Let us assume that at $t=0$ each subsystem is prepared in a thermal equilibrium at inverse temperature $\beta_i$ and is uncorrelated from the other sub-systems:
\begin{align}
\rho_0 = \bigotimes_{i=1}^N \rho_0^{i,\text{eq}}, \quad \rho_0^{i,\text{eq}}=\frac{e^{-\beta_i H_i(0)}}{Z_i(0)} 
\label{eq:eq-initial2}
\end{align}
with $Z_i(0)=\Tr_i e^{-\beta_i H_i(0)}$ being the partition functions. Let 
\begin{align}
\rho_t = U_t \rho_0 U_t^\dagger, \qquad \rho_t^i = \Tr_i \rho_t 
\end{align}
be the state of the full system and of each subsystem at time $t$, as resulting from the evolution $U_t$ generated by the full Hamiltonian $H(t)$. The energy dissipation in subsystem $i$ is given by:
\begin{align}
W_\text{diss}^i = \Tr_i [H_i(\tau)\rho_\tau^i -H_i(0)\rho_0^i] - \Delta F_i
\label{eq:Wdiss-i}
\end{align}
where 
\begin{align}
\Delta F_i = -\beta_i^{-1}\ln\frac{Z_i(\tau)}{Z_i(0)} 
\end{align}
Consider now the reference thermal equilibrium of subsystem $i$ corresponding to the value $H(t)$  of its Hamiltonian
\begin{align}
\rho_t^{i,\text{eq}} = \frac{e^{-\beta_i H_i(t)}}{Z_i(t)} 
\label{eq:rho-i-eq}
\end{align}
This is the state that subsystem $i$ would reach if at time $t$ one would turn off the coupling $V$, freeze the subsystem Hamiltonian to $H(t)$, weakly couple the system with a thermal bath at temperature $\beta_i$ and let the system relax to thermal equilibrium. Solving Eq. (\ref{eq:rho-i-eq}) for $\beta_i H_i(t)$ and plugging into (\ref{eq:Wdiss-i})
one finds
\begin{align}
\beta_i W_\text{diss}^i = -\Tr_i \rho_\tau^{i,\text{eq}} \ln \rho_\tau^{i} +\Tr_i \rho_0^{i,\text{eq}} \ln \rho_0^{i,\text{eq}}
\end{align}
adding and subtracting the quantity $\Tr_i \rho_\tau^{i,\text{eq}} \ln \rho_\tau^{i,\text{eq}}$, and summing over $i$ one obtains
\begin{align}
\sum_i \beta_i W_\text{diss}^i = \sum_i D[\rho^i_\tau || \rho_\tau^\text{i,eq}] + \sum_i \Delta S_i
\end{align}
where $\Delta S_i = S[\rho_\tau^\text{i}]-S[\rho_0^\text{i,\text{eq}}] $
denotes the change in von-Neumann information ($S[\sigma]=-\Tr \sigma \ln \sigma$ ) of sub-system $i$ from time $0$ to time $t$. Note 
that if the evolution taking $\rho_0^{i,\text{eq}}$  to $\rho_t^i$ were unitary, then the change in von Neumann information $\Delta S_i$ would be null. 
This happens if the system consists of a single subsystem (i.e. if $N=1$), in which case one recovers the known result in (\ref{eq:Wdiss}). For $N>1$ 
however the evolution of each reduced density matrix is evidently non-unitary, and this can in general lead to non null von Neumann information 
changes $\Delta S_i$. The last step of our argument consists in proving that  
\begin{align}
 \sum_i \Delta S_i = I[\rho_t] \doteq \sum_i S[\rho^i_\tau]- S[\rho_\tau]
\end{align}
which in turn implies Eq. (\ref{eq:sumbetaWdiss}).
To this end recall that the von Neumann information of the tensor product of several density matrices is the sum of the individual von Neumann informations. 
Therefore, since the initial state is factorised, Eq. (\ref{eq:eq-initial2}), then  $\sum_i S[\rho^{i,\text{eq}}_0]= S[\rho_0]$. But, due to the unitarity of the 
evolution of the full density matrix, it is $  S[\rho_0]=S[\rho_\tau]$ hence $ \sum_i \Delta S_i = \sum_i S[\rho^i_\tau]-\sum_i S[\rho^i_0]=\sum_i 
S[\rho^i_\tau]- S[\rho_\tau]$, that is $I[\rho_t]$. Note that the total correlation $I[\rho_\tau]$ is non-negative: the system starts in a factorised, i.e. uncorrelated 
state, and in the course of time, the evolution can only increase the correlation. The more the correlation established within the parts, the more the total dissipation. \footnote{We notice the interesting fact that the sum of lags and correlations can also be expressed as a lag in the full Hilbert space $ \sum_i D[\rho^i_t || \rho_t^\text{i,eq}] + I[\rho_t] = D[\rho_t || \otimes\rho_t^{i,\text{eq}}]$ (this is not true in general but depends on  the initial state being that of Eq. (\ref{eq:eq-initial2})), hence the multi-partite version of Eq. (\ref{eq:Wdiss}) can also be written $\sum \beta_i W_\text{diss}^i = D[\rho_t || \otimes\rho_t^{i,\text{eq}}]$, which is not immediately evident.}

\section{Application to heat engines}
Equation (\ref{eq:sumbetaWdiss}) has deep consequences in assessing the performance of thermal machines (large and small alike, classical or quantum). 
In this respect it is worth remarking that the whole argument presented above can be repeated within the formalism of classical Hamiltonian mechanics, 
with density matrices replaced by Liouville densities in phase space and traces replaced by phase space integrals. 
The crucial assumption  is that most thermal machines can be understood as multi-partite systems initially staying in an uncorrelated 
product of $N$ thermal states at different temperatures, and undergoing, as a whole, a unitary process. Consider for example the textbook 
Carnot engine. A device, which can range from a macroscopic gas in a box to a single two level system, is initially in thermal contact and equilibrium 
with a bath at temperature $T_1$, and is neither in contact nor correlated with a second bath at temperature $T_2$. The device plus bath at $T_1$ 
forms subsystem 1, with Hamiltonian $H_1$ initially staying at $\rho_t^\text{1,eq}=e^{-\beta_1 H_1}/Z_1$. The bath at temperature $T_2$ forms 
subsystem 2 with Hamiltonian $H_2$ and initially staying at $\rho_t^\text{2,eq}=e^{-\beta_2 H_2}/Z_2$. The initial state of the bipartite system is 
the factorised state in Eq. (\ref{eq:eq-initial2}). Next a number of manipulations acting on the system and on the coupling with the heat reservoirs, 
are performed on the total Hamiltonian. This process fits within the general scheme presented above. In general any other thermal machine that 
can be modelled as a driven open system in operating between various thermal baths can be modelled with our general scheme as well.

For a heat engine that works in a cycle between $N$ reservoirs, the final Hamiltonian $H(\tau)$ coincides with the initial Hamiltonian $H(0)$. In this case Eq. (\ref{eq:sumbetaWdiss}) simplifies into:
\begin{align}
\sum_i \beta_i \langle \Delta E_i \rangle = \sum_i D[\rho^i_\tau || \rho_0^\text{i,eq}] + I[\rho_t]
\label{eq:SumBetaiEi}
\end{align}
where $\langle \Delta E_i \rangle = \Tr_i H_i(\lambda_0^i)[\rho^i_\tau-\rho_0^\text{i,eq}]$ is the energy that remains stored in subsystem $i$ after the driving is over. Note that under the assumption that at the end of the driving the working substance stores no energy or a negligible amount thereof, and assuming all device-baths coupling are weak, $\langle \Delta E_i \rangle $ can be understood as the negative heat $-Q_i$ that has gone in the bath $i$. Also the sum $\sum_i \langle \Delta E_i \rangle$ quantifies the total energy injected $ \langle W \rangle$ into the full system, i.e., the negative work output
\begin{align}
 \langle W \rangle  =  \sum_i \langle \Delta E_i \rangle= - W_\text{out}
\label{eq:work}
\end{align}
Using Eqs. (\ref{eq:SumBetaiEi},\ref{eq:work}) and assuming the ordering $\beta_i\leq \beta_{i+1}$, we obtain Eq. (\ref{eq:complementarity}) where $\delta \eta = \eta^C/\eta-1$, with $\eta^C= 1-\beta_1/\beta_2$ the Carnot efficiency and $\eta= \langle W \rangle /  \langle \Delta E_1 \rangle= W_\text{out}/(-Q_1)$, the engine efficiency.
In the specific case when $N=2 $, the inequality in (\ref{eq:complementarity}) reduces to an equality
\begin{align}
\beta_2 W_\text{out} \delta \eta =  D[\rho^1_\tau || \rho_0^\text{1,eq}]+ D[\rho^2_\tau || \rho_0^\text{2,eq}] +  I[\rho_\tau]
\label{eq:complementarity2}
\end{align}

Assuming that the machine operates as a  heat engine (i.e., with our sign conventions, $W_\text{out}\geq 0$),  since the r.h.s. of Eq. (\ref{eq:complementarity}) 
is non-negative it implies $\eta \leq \eta^C$. That is no heat engine can overcome the Carnot efficiency. A similar conclusion can be drawn for refrigerators.

Eq.(\ref{eq:complementarity}) has however much deeper and stronger implications. Can Carnot efficiency coexist with finite work output? 
At Carnot efficiency the left hand side of Eq. (\ref{eq:complementarity}) is null. Since the right hand side is the sum of three non-negative terms, 
each of them must vanish at Carnot efficiency. Recall that the relative entropy $D[\cdot || \cdot]$ only vanishes when its two arguments are 
identical. This implies that at Carnot efficiency $\rho^i_\tau = \rho_0^\text{i,eq}$. Hence $\langle \Delta E_i \rangle = 0 $ and  $\sum_i \langle \Delta E_i 
\rangle = \langle W\rangle = 0$. This implies a very strong result: Any machine working exactly at Carnot efficiency delivers/absorbs no heat and no work. 
This seems to contrast with the textbook knowledge that the Carnot engine delivers a finite work (given by the non-null area enclosed by the cycle in 
the $S-T$ plane). To reconcile these two apparently incompatible facts notice that we have modelled the evolution of the engine as a global unitary $U$ 
that involves both working substance and baths, and changes not only the state of the working substance but that of the bath as well. If you assume 
the latter to have an infinite heat capacity so as to be resilient to the external action, as is implied in the textbook treatment, the work is then indeed 
infinitesimally small when compared to the infinite energy stored in the bath. In a separate publication \ref{...} we have addressed how 
one can asymptotically approach the Carnot point at finite power per number of constituents of the working substance.

\section{Two-qubit engine}
\begin{figure}[t]
\includegraphics[width=0.8\linewidth]{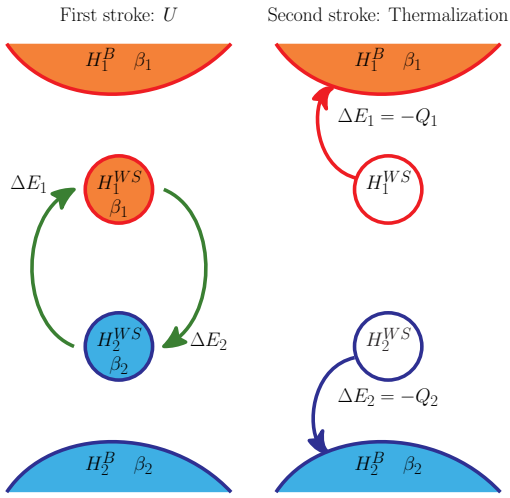}
\caption{Schematics of the two-qubits engine \cite{Campisi15NJP17}. First stroke: A unitary transformation leads the two qubits out-of equilibrium. 
Second stroke: thermalisation restores the initial bi-partite equilibrium, Eq. (\ref{eq:eq-initial1}). }
\label{fig:2}
\end{figure}

We consider the experimentally feasible two-qubit engine described in Ref.\cite{Campisi15NJP17} and first introduced in~\cite{Quan07PRE76}, 
see Fig. \ref{fig:2}. Two uncoupled qubits are prepared each in thermal equilibrium with a distinct reservoir. Between time $t=0$ and $t=\tau$
the two qubits are coupled to each other by means of coupling $V(t)$ and the coupling with their respective reservoirs is sufficiently weak that we 
can assume they evolve unitarily.  Hence we can apply the theory developed above to the small Hilbert space of the two qubits, rather than the total 
qubits plus reservoirs. This allows for a drastic simplification. The Hamiltonian of the two qubits reads:
\begin{align}
H(t) = H_1+H_2+ V(t)=\frac{\omega_1}{2}\sigma_z^1 + \frac{\omega_2}{2}\sigma_z^2 + V(t)
\end{align}
with $\sigma_z^i$ the Pauli matrices.
After the coupling is applied each qubit is in an out of equilibrium state, and is then allowed to relax to thermal equilibrium with its own thermal bath  
for a sufficient time $\tau_\text{relax}$ so that the initial state is restored and the cycle is completed. The energies $\langle \Delta E_i \rangle $ 
delivered to each qubit during the drive equal the negative heat $Q_i$ ceded to their respective reservoir during the relaxation step.
\begin{figure}[t]
\includegraphics[width=\linewidth]{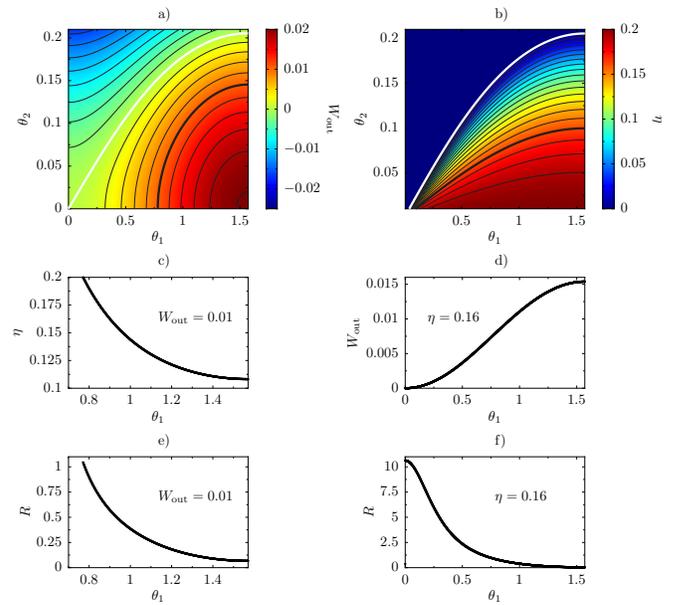}
\caption{Work output, efficiency, and relative correlation strength $R$, Eq. (\ref{eq:R}), in a partial SWAP-CNOT two-qubit engine, Eq. (\ref{eq:SwapCnot}). 
Panel a) Work output as function of $\theta_1,\theta_2$. White curve is $W_\text{out}=0$.  Thick black curve is $W_\text{out}=0.01$.
Panel b) Efficiency as function of $\theta_1,\theta_2$. White curve is $\eta=0$. Thick black curve is $\eta=0.16$. 
Panel c) Efficiency at fixed $W_\text{out}=0.01$.
Panel d) Work output at fixed $\eta=0.16$. 
Panel e) Relative correlation strength, Eq. (\ref{eq:R}), at fixed $W_\text{out}=0.01$.
Panel f) Relative correlation strength, Eq. (\ref{eq:R}), at fixed $\eta=0.16$. 
Here $\beta_1/\beta_2=0.5$ and $\omega_2/\omega_1=0.8$.}
\label{fig:3}
\end{figure}
It is known \cite{Campisi15NJP17} that heat engine operation is achieved for $\beta_1/\beta_2 \leq \omega_2/\omega_1 \leq 1$ irrespective of the temperatures of the two reservoirs and that the maximal efficiency that can be achieved is $\eta = 1- \omega_2/\omega_1$ and . The maximal efficiency is achieved by any partial swap operation, that is a rotation in the subspace spanned by the states $| +-  \rangle$, 
$|- + \rangle$. Among all partial swaps, full swaps (that is operations that map $| +-  \rangle$ into $|- + \rangle$ (modulo a phase) and vice-versa) achieve maximal work output.
This behaviour can now be understood in terms of Eq. (\ref{eq:complementarity}): among all partial swaps, full swaps produce no correlation.
It is also known \cite{Campisi15NJP17} that at Carnot efficiency (i.e., when $\omega_2/\omega_1 \rightarrow \beta_1/\beta_2 $) the work output vanishes. This is now understood on the basis of Eq. (\ref{eq:complementarity2}). It is worth emphasising that correlation and lags are not independent on each other because they are both functions of $\rho_\tau$. This gives rise to a rich interplay between the two sources of dissipation. To get an insight into this interplay we consider the following gate operation:
\begin{align}
U = U_\text{SWAP}(\theta_1) \cdot U_\text{CNOT}(\theta_2)
\label{eq:SwapCnot}
\end{align}
where $U_\text{CNOT}(\theta_2)$ is a partial CNOT gate, that is a rotation of an angle $\theta_2$ in the space spanned by $| -+  \rangle$, $| - -  \rangle$, followed by a partial swap  $U_\text{SWAP}(\theta_1)$, that is a rotation in the space spanned by $| -+  \rangle$, $| + -  \rangle$.
Fig. \ref{fig:3} presents contour plots of work and efficiency as a function of $\theta_1$ and $\theta_2$. It also shows how the ratio 
\begin{align}
R[\rho_\tau] \doteq \frac{ I[\rho_\tau]}{D[\rho^1_\tau || \rho_0^\text{1,eq}]+ D[\rho^2_\tau || \rho_0^\text{2,eq}]}
\label{eq:R}
\end{align}
of correlation over lags behave at fixed work and efficiency. The plots highlight how at fixed efficiency the lesser the relative strength of correlations, as compared to lags, the more the work output. Vice-versa, at fixed work output, the higher the relative strength of correlations the higher the efficiency. This evidences the very remarkable fact that the build-up of correlations do not necessarily represent a thermodynamic cost. Whether it does so may depend on whether one is interested in the work output or the efficiency. 
By comparison of the two contour plots it is also possible to see that not necessarily an increase in work output is accompanied by a decrease in efficiency. It is in fact possible to draw curves in the $(\theta_1,\theta_2)$ plane where they both increase (not shown). A similar occurrence has been recently noticed in \cite{2016arXiv160207016R}.

\section{Conclusions}
We have identified correlations and lags as the two sources of dissipation in driven multipartite systems, (Eq. (\ref{eq:complementarity}). In the case of 
heat engines, this implies the inequality (\ref{eq:complementarity}), concerning the product of deviation from Carnot's efficiency and the work output. 
The interplay of lags and correlations is illustrated with a two-qubit engine driven by a partial CNOT-SWAP gate. The example shows that correlation 
build up is not necessarily a thermodynamic cost. It can depend on whether one looks at work or efficiency.

\subsection{Acknowledgements}
This research was supported by the 7th European Community Framework Programme under grant agreements n. 623085 (MC-IEF-NeQuFlux), 
n. 600645 (IP-SIQS), n. 618074 (STREP-TERMIQ) and by the COST action MP1209 ``Thermodynamics in the quantum regime''.

\end{document}